\documentclass[journal,comsoc]{IEEEtran}
\usepackage[T1]{fontenc}

\ifCLASSINFOpdf
  \usepackage[pdftex]{graphicx}
  \graphicspath{{./png/}{./pdf/}{./jpeg/}}
  \DeclareGraphicsExtensions{.pdf,.jpeg,.png}
\else
\fi
\usepackage{amsmath}
\usepackage[cmintegrals]{newtxmath}
\ifCLASSOPTIONcompsoc
  \usepackage[caption=false,font=normalsize,labelfont=sf,textfont=sf]{subfig}
\else
  \usepackage[caption=false,font=footnotesize]{subfig}
\fi

\usepackage{stfloats}
\usepackage{url}

\usepackage{todonotes}
\usepackage[detect-weight=true, detect-family=true, per-mode=symbol]{siunitx}
\DeclareSIUnit{\bits}{bits}
\usepackage{booktabs}
\renewcommand{\vec}[1]{\ensuremath{\boldsymbol{#1}}}

\usepackage[acronym, nomain,nogroupskip,nonumberlist,nopostdot, toc]{glossaries}
\newacronym{3GPP}{3GPP}{3rd Generation Partnership Project }
\newacronym{4G}{4G}{fourth generation}
\newacronym{5G}{5G}{fifth generation}
\newacronym{ADC}{ADC}{analogue-to-digital converter}
\newacronym{ADS}{ADS}{advanced design system}
\newacronym{AGC}{AGC}{automatic gain control}
\newacronym{AoA}{AoA}{angle of arrival}
\newacronym{AoD}{AoD}{angle of departure}
\newacronym{AP}{AP}{Access Point}
\newacronym{AR}{AR}{angular resolution}
\newacronym{AS}{AS}{angular selectivity}
\newacronym{AWGN}{AWGN}{Additive White Gaussian Noise}
\newacronym{BBU}{BBU}{Base Band Unit}
\newacronym{BCC}{BCC}{Bristol City Council}
\newacronym{BER}{BER}{bit error rate}
\newacronym{BF}{BF}{beamforming}
\newacronym{BIO}{BIO}{Bristol Is Open}
\newacronym{BS}{BS}{base station}
\newacronym{BT}{BT}{British Telecom}
\newacronym{CCDF}{CCDF}{complementary cumulative distribution function}
\newacronym{CDF}{CDF}{cumulative distribution function}
\newacronym{CDMA}{CDMA}{Code Division Multiple Access}
\newacronym{CDT}{CDT}{Centre for Doctoral Training}
\newacronym{CFO}{CFO}{carrier frequency offset}
\newacronym{CNC}{CNC}{computer numerical control}
\newacronym{COTS}{COTS}{commercial off-the-shelf}
\newacronym{CP}{CP}{Cyclic Prefix}
\newacronym{CPA}{CPA}{coordinated pilot assignment}
\newacronym{CRC}{CRC}{cyclic redundancy check}
\newacronym{CSI}{CSI}{channel state information}
\newacronym{CSIRO}{CSIRO}{Commonwealth Scientific and Industrial Research Organisation}
\newacronym{CSN}{CSN}{Communication Systems \& Networks}
\newacronym{D2D}{D2D}{device-to-device}
\newacronym{DAC}{DAC}{digital-to-analogue converter}
\newacronym{dB}{dB}{decibels}
\newacronym{dBm}{dBm}{decibel-milliwatts}
\newacronym{DC}{DC}{direct current}
\newacronym{DL}{DL}{downlink}
\newacronym{DMA}{DMA}{direct memory access}
\newacronym{DMRS}{DMRS}{demodulation reference signal}
\newacronym{DRAM}{DRAM}{dynamic random access memory}
\newacronym{DS}{DS}{delay-spread}
\newacronym{DSP}{DSP}{digital signal processing}
\newacronym{EM}{EM}{electro-magnetic}
\newacronym{EPSRC}{EPSRC}{Engineering and Physical Sciences Research Council}
\newacronym{ESD}{ESD}{electro-static discharge}
\newacronym{EVM}{EVM}{error vector magnitude}
\newacronym{FFT}{FFT}{fast Fourier transform}
\newacronym{FDD}{FDD}{frequency division duplex}
\newacronym{FD-MIMO}{FD-MIMO}{full dimension \gls{MIMO}}
\newacronym{FEC}{FEC}{forward-error correction}
\newacronym{FIFO}{FIFO}{first-in first-out}
\newacronym{FITRA}{FITRA}{fast iterative truncation algorithm}
\newacronym{FlexRIO}{FlexRIO}{flexible \gls{RIO}}
\newacronym{FOC}{FOC}{frequency offset correction}
\newacronym{FPGA}{FPGA}{field-programmable gate array}
\newacronym{FWA}{FWA}{fixed wireless access}
\newacronym{FXP}{FXP}{fixed-point}
\newacronym{GPS}{GPS}{global positioning system}
\newacronym{GPSDO}{GPSDO}{\gls{GPS} disciplined oscillator}
\newacronym{GNSS}{GNSS}{Global Navigation Satellite Systems}
\newacronym{GUI}{GUI}{graphical user interface}
\newacronym{H2T}{H2T}{Host-to-target}
\newacronym{HD}{HD}{High-Definition}
\newacronym{HDL}{HDL}{Hardware Description Language}
\newacronym{HPE}{HPE}{Hewlett Packard Enterprise}
\newacronym{HPBW}{HPBW}{half power beam width}
\newacronym{HQ}{HQ}{headquarters}
\newacronym{ICI}{ICI}{inter-carrier interference}
\newacronym{ICT}{ICT}{Information and Communications Technology}
\newacronym{ID}{ID}{identity}
\newacronym{IDE}{IDE}{integrated development environment}
\newacronym{IFFT}{IFFT}{Inverse Fast Fourier Transform}
\newacronym{iid}{iid}{independent and indentically distributed}
\newacronym{IO}{IO}{Input/Output}
\newacronym{IoT}{IoT}{internet of things}
\newacronym{IP}{IP}{Intellectual Property}
\newacronym{ISI}{ISI}{Inter-Symbol Interference}
\newacronym{IUI}{IUI}{inter-user interference}
\newacronym{ISM}{ISM}{industrial, scientific and measurement}
\newacronym{JSDM}{JSDM}{Joint Spatial Division Multiplexing}
\newacronym{KPH}{KPH}{kilometres per hour}
\newacronym{KPI}{KPI}{key performance indicator}
\newacronym{LDPC}{LDPC}{low-density parity-check}
\newacronym{LED}{LED}{Light Emitting Diode}
\newacronym{LF}{LF}{Loading Factor}
\newacronym{LIDAR}{LIDAR}{Laser Illuminated Detection \& Ranging}
\newacronym{LO}{LO}{local oscillator}
\newacronym{LOS}{LOS}{line-of-sight}
\newacronym{LS}{LS}{least squares}
\newacronym{LSD}{LSD}{Local Search Detection}
\newacronym{LTE}{LTE}{long-term evolution}
\newacronym{LUT}{LUT}{Look-up Table}
\newacronym{M2M}{M2M}{machine-to-machine}
\newacronym{MAC}{MAC}{Medium Access Control}
\newacronym{MMIMO}{MMIMO}{massive \gls{MIMO}}
\newacronym{MF}{MF}{matched filtering}
\newacronym{MRT}{MRT}{maximal ratio transmission}
\newacronym{MCS}{MCS}{modulation and coding scheme}
\newacronym{MGS}{MGS}{modified gram schmidt}
\newacronym{MIMO}{MIMO}{multiple-input, multiple-output}
\newacronym{ML}{ML}{Maximum likelihood}
\newacronym{MSE}{MSE}{Mean Square Error}
\newacronym{MMSE}{MMSE}{minimum mean square error}
\newacronym{MPD}{MPD}{Message Passing Detection}
\newacronym{MRC}{MRC}{maximal ratio combining}
\newacronym{MU}{MU}{multi-user}
\newacronym{MUI}{MUI}{Multi-User Interference}
\newacronym{MUSIC}{MUSIC}{Multiple Signal Classification}
\newacronym{MWC}{MWC}{mobile world congress}
\newacronym{MXI}{MXI}{Multisystem Extension Interface}
\newacronym{NF}{NF}{noise figure}
\newacronym{NI}{NI}{National Instruments}
\newacronym{NLOS}{NLOS}{non-line-of-sight}
\newacronym{OBR}{OBR}{Out-of-band power ratio}
\newacronym{OCXO}{OCXO}{Oven-controlled Crystal Oscillator}
\newacronym{OFDM}{OFDM}{orthogonal frequency division multiplexing}
\newacronym{OTA}{OTA}{over-the-air}
\newacronym{PA}{PA}{Power Amplifier}
\newacronym{PAPR}{PAPR}{peak-to-average power ratio}
\newacronym{PAS}{PAS}{Power-Azimuth Spectrum}
\newacronym{PC}{PC}{Power Control}
\newacronym{PCIe}{PCIe}{peripheral component interconnect express}
\newacronym{PDP}{PDP}{Power Delay Profile}
\newacronym{PDSCH}{PDSCH}{Physical Downlink Shared Channel}
\newacronym{PHY}{PHY}{physical layer}
\newacronym{PL}{PL}{Path-Loss}
\newacronym{PLL}{PLL}{Phase-locked Loop}
\newacronym{PMP}{PMP}{precoding, modulation and \gls{PAPR} reduction}
\newacronym{PN}{PN}{pseudo-random}
\newacronym{POC}{POC}{proof-of-concept}
\newacronym{P2P}{P2P}{Peer-to-peer}
\newacronym{ppb}{ppb}{parts-per-billion}
\newacronym{PPC}{PPC}{Pilot Power Control}
\newacronym{PSS}{PSS}{primary synchronisation signal}
\newacronym{PTS}{PTS}{Pilot Time Slot}
\newacronym{PUSCH}{PUSCH}{Physical Uplink Shared Channel}
\newacronym{PXIe}{PXIe}{\gls{PCIe} eXtensions for instrumentation}
\newacronym{QAM}{QAM}{quadrature amplitude modulation}
\newacronym{QPSK}{QPSK}{quadrature phase-shift keying}
\newacronym{QRD}{QRD}{QR decomposition}
\newacronym{RAN}{RAN}{Radio Access Network}
\newacronym{RB}{RB}{resource block}
\newacronym{R&D}{R\&D}{Research and development}
\newacronym{RE}{RE}{Resource Element}
\newacronym{RF}{RF}{radio frequency}
\newacronym{RIO}{RIO}{reconfigurable input/output}
\newacronym{RMS}{R.M.S.}{root-mean-squre}
\newacronym{RRH}{RRH}{remote radio head}
\newacronym{RSS}{RSS}{Received Signal Strength}
\newacronym{RT}{RT}{Ray-tracing}
\newacronym{RTM}{RTM}{Rear Transition Module}
\newacronym{Rx}{Rx}{Receive}
\newacronym{SC}{SC}{Single-Carrier}
\newacronym{SDMA}{SDMA}{Space-Divison Multiple Access}
\newacronym{SDR}{SDR}{software-defined radio}
\newacronym{SE}{SE}{spectral efficiency}
\newacronym{SER}{SER}{Symbol Error Rate}
\newacronym{SINR}{SINR}{signal to interference plus noise ratio}
\newacronym{SIR}{SIR}{signal to interference ratio}
\newacronym{SM}{SM}{Spatial Modulation}
\newacronym{SMA}{SMA}{subminiature version A}
\newacronym{SNR}{SNR}{signal to noise ratio}
\newacronym{SPDT}{SPDT}{Single Pole Double Throw}
\newacronym{SSD}{SSD}{solid-state drive}
\newacronym{SISO}{SISO}{Single-Input Single-Output}
\newacronym{STBC}{STBC}{Space-Time Block-Coding}
\newacronym{SVD}{SVD}{singular value decomposition}
\newacronym{SVS}{SVS}{singular value spread}
\newacronym{T2H}{T2H}{Target-to-host}
\newacronym{TA}{TA}{timing advance}
\newacronym{TCF}{TCF}{temporal correlation function}
\newacronym{TDD}{TDD}{time division duplex}
\newacronym{TDOA}{TDOA}{Time-Difference-of-Arrival}
\newacronym{TM}{TM}{transmission mode}
\newacronym{TO}{TO}{Timing Offset}
\newacronym{TPC}{TPC}{Transmitter Power Control}
\newacronym{TR}{TR}{tone-reservation}
\newacronym{Tx}{Tx}{Transmit}
\newacronym{TTI}{TTI}{Transmission Time Interval}
\newacronym{U8}{U8}{Unsigned 8-bit}
\newacronym{UDN}{UDN}{ultra-dense network}
\newacronym{UDP}{UDP}{User Datagram Protocol}
\newacronym{UE}{UE}{user equipment}
\newacronym{UI}{UI}{User Interface}
\newacronym{UL}{UL}{uplink}
\newacronym{ULA}{ULA}{uniform linear array}
\newacronym{USRP}{USRP}{universal software radio peripheral}
\newacronym{UoB}{UoB}{University of Bristol}
\newacronym{V2I}{V2I}{vehicle-to-infrastructure}
\newacronym{V2V}{V2V}{vehicle-to-vehicle}
\newacronym{VLC}{VLC}{VideoLAN Client}
\newacronym{VTC}{VTC}{Vehicular Technology Conference}
\newacronym{WARP}{WARP}{Wireless Open-Access Research Platform}
\newacronym{ZF}{ZF}{zero-forcing}
\begin{document}
\bstctlcite{IEEEexample:BSTcontrol}
%
\title{Achievable Rates and Training Overheads for a Measured LOS Massive MIMO Channel}
%
%
%

\author{Paul~Harris,
		Wael~Boukley~Hasan,
		Liang~Liu,
		Steffen~Malkowsky,
		Mark~Beach,
        Simon~Armour,
        Fredrik~Tufvesson
        and Ove~Edfors
        
\thanks{P. Harris, Wael Boukley Hasan, M.A. Beach and S. Armour are with the \gls{CSN} Group at the University of Bristol, U.K. Email: \{paul.harris, wb14488, m.a.beach, simon.armour\}@bristol.ac.uk
	
S. Malkowsky, O. Edfors, L. Liu and F. Tufvesson are with the Dept. of Electrical and Information Technology, Lund University, Sweden. Email: \{firstname.lastname\}@eit.lth.se}
}

\maketitle

\begin{abstract}
This paper presents achievable \gls{UL} sum-rate predictions for a measured \gls{LOS} \gls{MMIMO} scenario and illustrates the trade-off between spatial multiplexing performance and channel de-coherence rate for an increasing number of \gls{BS} antennas. In addition, an \gls{OFDM} case study is formed which considers the \SI{90}{\percent} coherence time to evaluate the impact of \gls{MMIMO} channel training overheads in high-speed \gls{LOS} scenarios. It is shown that whilst \SI{25}{\percent} of the achievable \gls{ZF} sum-rate is lost when the resounding interval is increased by a factor of \num{4}, the \gls{OFDM} training overheads for a \num{100}-antenna \gls{MMIMO} \gls{BS} using an LTE-like physical layer could be as low as \SI{2}{\percent} for a terminal speed of \SI{90}{\metre\per\second}.
\end{abstract}


\IEEEpeerreviewmaketitle

\section{Introduction}
\glsresetall
\IEEEPARstart{S}{ince} the conception of \gls{MMIMO} in~\cite{Marzetta2010}, one aspect that has remained of interest is the operation of \gls{MMIMO} under mobile conditions. Whilst the precision of decoding and precoding increases with a larger number of antennas, thereby improving spatial multiplexing performance, some techniques such as \gls{ZF} are more sensitive to \gls{CSI} error. The work in~\cite{DBLP:journals/corr/abs-1305-6151} analyses the impact of channel aging in \gls{MMIMO} using the Jakes model and highlights the possibility of partial mitigation using channel prediction. It is also shown in~\cite{7510708} through simulations that user rates for \gls{MMIMO} can drop radically for speeds as low as \SI{10}{\kilo\metre\per\hour} within urban environments. Similar results are presented for \gls{NLOS} scenarios in~\cite{7869082} using measured data, but it is also shown that outdoor \gls{LOS} conditions with minimal scattering can exhibit very high levels of stability. In strong \gls{LOS} conditions with minimal scattering, a Doppler shift will occur between a moving terminal and static \gls{BS} but the spread is minimal. Thus, \gls{OFDM} systems will experience reduced \gls{ICI} in these scenarios, even at high velocity, assuming the transmitters and receivers used can manage large \glspl{CFO}. Furthermore, the increased coherence time that results from a low Doppler spread in \gls{LOS} can allow more \gls{MMIMO} terminals to be trained for the same level of mobility, or an existing number to be supported at higher levels of mobility.

This paper serves as a companion paper to the work presented in~\cite{PaulJSAC}. Using data from the same measurement campaign, additional metrics are presented here to provide further insights on \gls{LOS} \gls{MMIMO} performance with terminal mobility. The authors believe these contributions will aid \gls{MMIMO} system design and potentially serve to validate modeling assumptions. The remainder of the paper is structured as follows. In Section~\ref{sec:scen}, a brief recap of the \gls{MMIMO} system and experiment configurations from~\cite{PaulJSAC} is given. The approaches for obtaining both the achievable \gls{UL} \gls{MMIMO} sum-rate and the expected performance under channel aging are then presented in Section~\ref{sec:cap}. Section~\ref{sec:res} provides results for the sum-rate, depicts the trade-off between spatial multiplexing performance and coherence distance as the number of \gls{BS} antennas $M$ is increased, shows the impact of channel aging on detector performance and considers an application of the measured coherence time to an \gls{OFDM} case study. The paper concludes in Section~\ref{sec:conc}.

\section{Measurement Scenario}
\label{sec:scen}
The results presented here use data acquired from the measurement campaign described in~\cite{PaulJSAC} where a \num{100}-antenna real-time \gls{MMIMO} \gls{BS} is used to serve \num{8} user terminals in a \gls{LOS} environment. The expected pilot \gls{SNR} is in the range of \SIrange{20}{30}{\decibel}. All sum-rate results presented were obtained using the full \SI{30}{\second} mobile scenario, whereas the evaluation of coherence time is based upon the movement of car \num{2} through the \SI{4}{\second} scenario subset shown in Fig. \num{9} of~\cite{PaulJSAC}. For the latter it is ensured that the vehicle speed does not exceed \SI{29}{\kilo\metre\per\hour} such that the channel capture rate of \SI{200}{\hertz} (\SI{5}{\milli\second} sounding period) meets the spatial Nyquist constraint of $\lambda/2$. For further details on the system or experiment, the reader should refer to the aforementioned paper.

\section{Capacity Evaluation}
\label{sec:cap}
The aggregate uplink sum-rate is computed here for a given resource block $b$, time instance $t$ and noise power $N$ as
\begin{equation}
\label{equ:Cap}
C_{\Delta}(b,t)=\sum_{i=1}^{K}\log_{2}\left(1+\frac{\left|\vec{h}_{i,b}^{norm}(t)\ \vec{w}_{i,b}^{T}(t-\Delta)\right|^{2}}{\sum_{j\neq i}\left|\vec{h}_{i,b}^{norm}(t)\ \vec{w}_{j,b}^{T}(t-\Delta)\right|^{2} +\ N}\right)
\end{equation}
where $\vec{h}_{i,b}^{norm}$ and $\vec{w}_{i,b}$ represent the channel and linear decoding vectors respectively for user $i$ and frequency resource block $b$. $\vec{w}_{i,b}$ is derived using either \gls{ZF} or \gls{MF} and $\Delta$ represents the time delay between \gls{CSI} estimation and the application of the respective decoder weights. The calculation is thus based upon the  \gls{AWGN} capacity formula of~\cite{6773067} and a similar approach was used in~\cite{7869082}. To avoid noise power correlation and sum-rate inaccuracies, the decoding matrix $\vec{W}_{b}$ is derived using the first \gls{CSI} estimate and applied to the second when $\Delta=0$. Applying a decoding matrix to the same channel matrix it was derived from is also impossible in a real system due to the fact that processing is not instantaneous and the \gls{CSI} estimation will contain errors. For \gls{ZF}, $\vec{W}_{b}=\left(\vec{H}_{b}\vec{H}_{b}^H\right)^{-1}\vec{H}_{b}^H$, and for \gls{MF}, $\vec{W}_{b}=M^{-1}\vec{H}_{b}^H$, where the latter division by the antenna dimension ensures the decoded \gls{MF} power for a single user (no interference) is equal to \num{1} and matches \gls{ZF}. The noise power $N$ is then chosen to provide the desired average \gls{SNR} for all users. In the cases where spectral efficiency is given per user, the median sum-rate across $K$ active user vectors is taken unless otherwise specified.

A statistical measure was also calculated to evaluate the percentage of the achievable sum-rate one can expect to obtain with ageing decoder weights. Referred to as the expected sum-rate, it is defined as shown in (\ref{eq:EC}) as the ratio of the achievable sum-rate given some channel resounding interval $\Delta$ to the achievable sum-rate when $\Delta=0$.
\begin{equation}
\label{eq:EC}
\gamma(\Delta) = \mathbf{E}\left\lbrace\frac{C_{\Delta}(t)}{C(t)}\right\rbrace
\end{equation}

\begin{figure}[!t]
	\centering
	\includegraphics[width=\columnwidth]{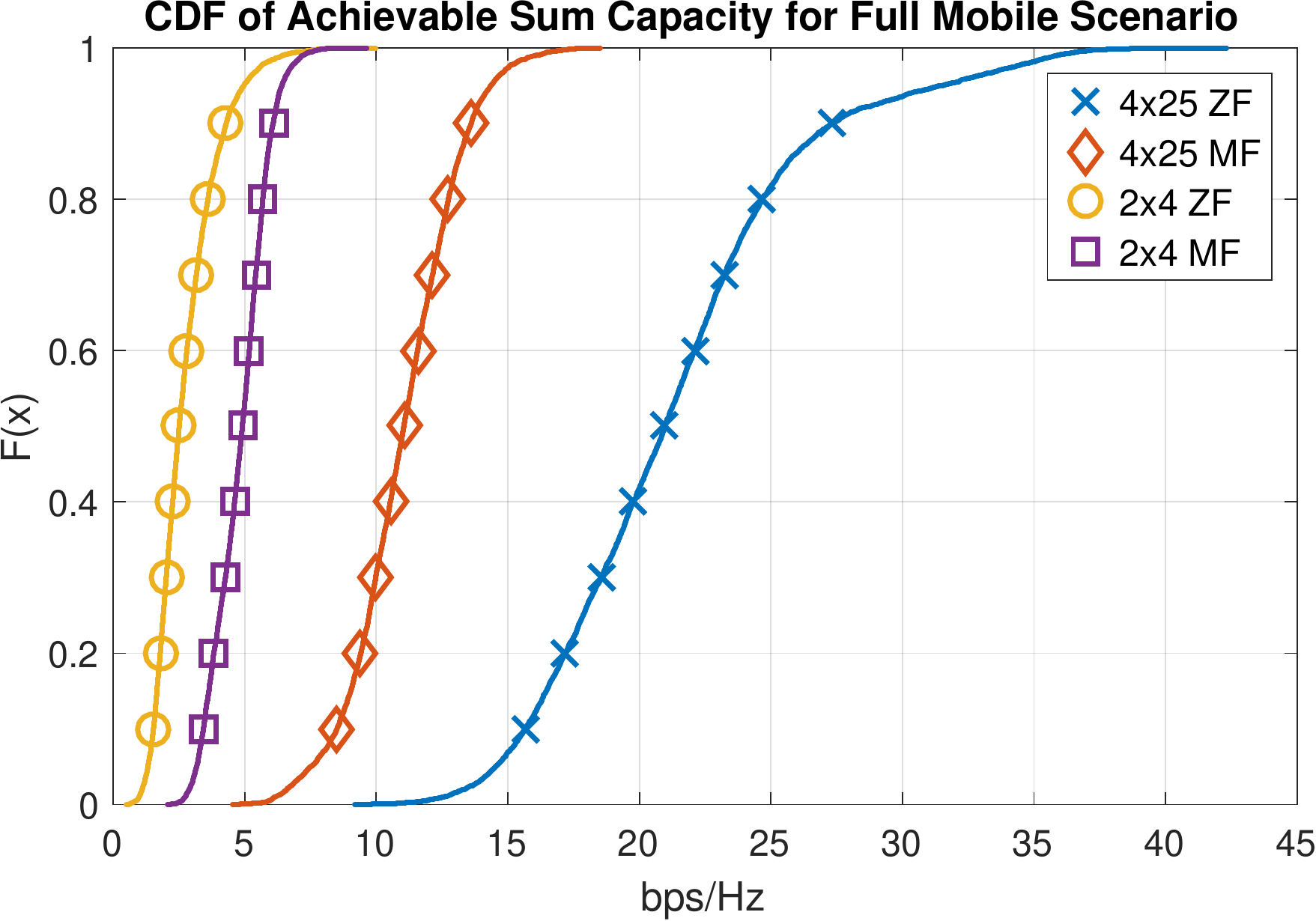}
	\caption{Achievable sum-rate CDF for the full mobile scenario using \gls{ZF} and \gls{MF} for \num{100} and \num{8} \gls{BS} antennas serving \num{4} dual-antenna \glspl{USRP} clients (\num{8} streams). Average \gls{SNR} fixed to \SI{30}{\decibel}}
	\label{fig:CapCDF}
\end{figure}
\section{Results}
\label{sec:res}
\subsection{Achievable Rate}
\begin{figure}[!t]
	\centering
	\includegraphics[width=\columnwidth]{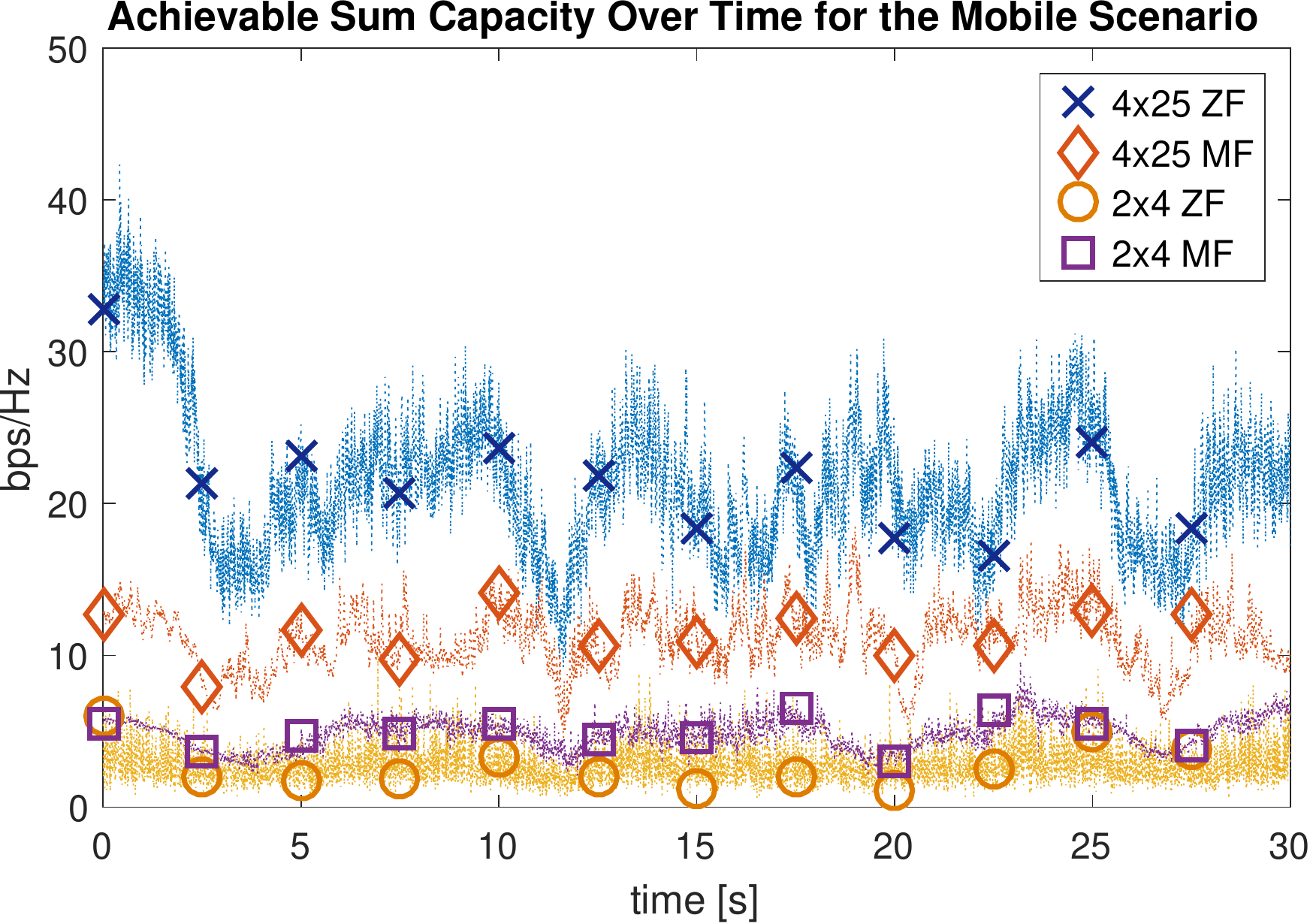}
	\caption{Achievable sum-rate over time for the full mobile scenario using \gls{ZF} and \gls{MF} for \num{100} and \num{8} \gls{BS} antennas serving \num{4} dual-antenna \glspl{USRP} clients (\num{8} streams). Average \gls{SNR} fixed to \SI{30}{\decibel}}
	\label{fig:CapTime}
\end{figure}

As described in Section~\ref{sec:cap}, the decoder matrices are calculated using the channel matrix one sample (\SI{5}{\milli\second}) prior to the channel matrix they are applied to in order to avoid noise summation and unrealistic sum-rate results.~\figurename~\ref{fig:CapCDF} and~\figurename~\ref{fig:CapTime} show both the achievable sum-rate \gls{CDF} and the results across time respectively over the full \num{30} second mobile scenario for the \num{4x25} and \num{2x4} \gls{BS} array configurations serving \num{4} dual-antenna \glspl{USRP} clients (\num{8} streams). The average \gls{SNR} was fixed to \SI{30}{\decibel}. With the full \num{100} antenna configuration, \gls{ZF} offers the best performance, providing a median achievable spectral efficiency of \SI{21}{\bits\per\second\per\hertz} and peaks in the \SIrange{30}{40}{\bits\per\second\per\hertz} range. The median \gls{MF} performance is a factor of two smaller at \SI{11}{\bits\per\second\per\hertz}, but exhibits less variance, spanning \SIrange{5}{18}{\bits\per\second\per\hertz} compared to \SIrange{9}{43}{\bits\per\second\per\hertz} for \gls{ZF}. When the time domain representation is inspected (~\figurename~\ref{fig:CapTime}), it is possible to see the impact of terminal movement upon the achievable sum-rate. At the beginning of the scenario when the pedestrians and cars start to move, there is an obvious impact for the \num{100} antenna \gls{ZF} case. Whilst static, the achievable sum-rate ranges between \SIrange{30}{40}{\bits\per\second\per\hertz}, but after two seconds once movement has commenced, the sum-rate drops sharply by half and averages \SI{20}{\bits\per\second\per\hertz} for the remainder of the scenario. \gls{MF}, whilst performing worse, experiences far less variance across the scenario, indicating the impact of channel aging is not as severe. In the \num{2x4} case, the \gls{ZF} performance is very poor. The variance it exhibits around a mean value of \SI{2.7}{\bits\per\second\per\hertz} implies the equalization resulted in very low \gls{SINR}. Across all time samples and all users, the mean \gls{SINR} following equalization using \gls{ZF} was \SI{-6.8}{\decibel}, compared to \SI{-2.8}{\decibel} for \gls{MF}. In such a case it is better to yield the full coherent beamforming gain of $M$ provided by \gls{MF} and tolerate the interference, but the result is still not ideal, and the benefits of large $M$ for linear detection can clearly be seen.

\subsection{The Impact of Antennas on CSI Update Rates}
\begin{figure}[!t]
	\centering
	\includegraphics[width=\columnwidth]{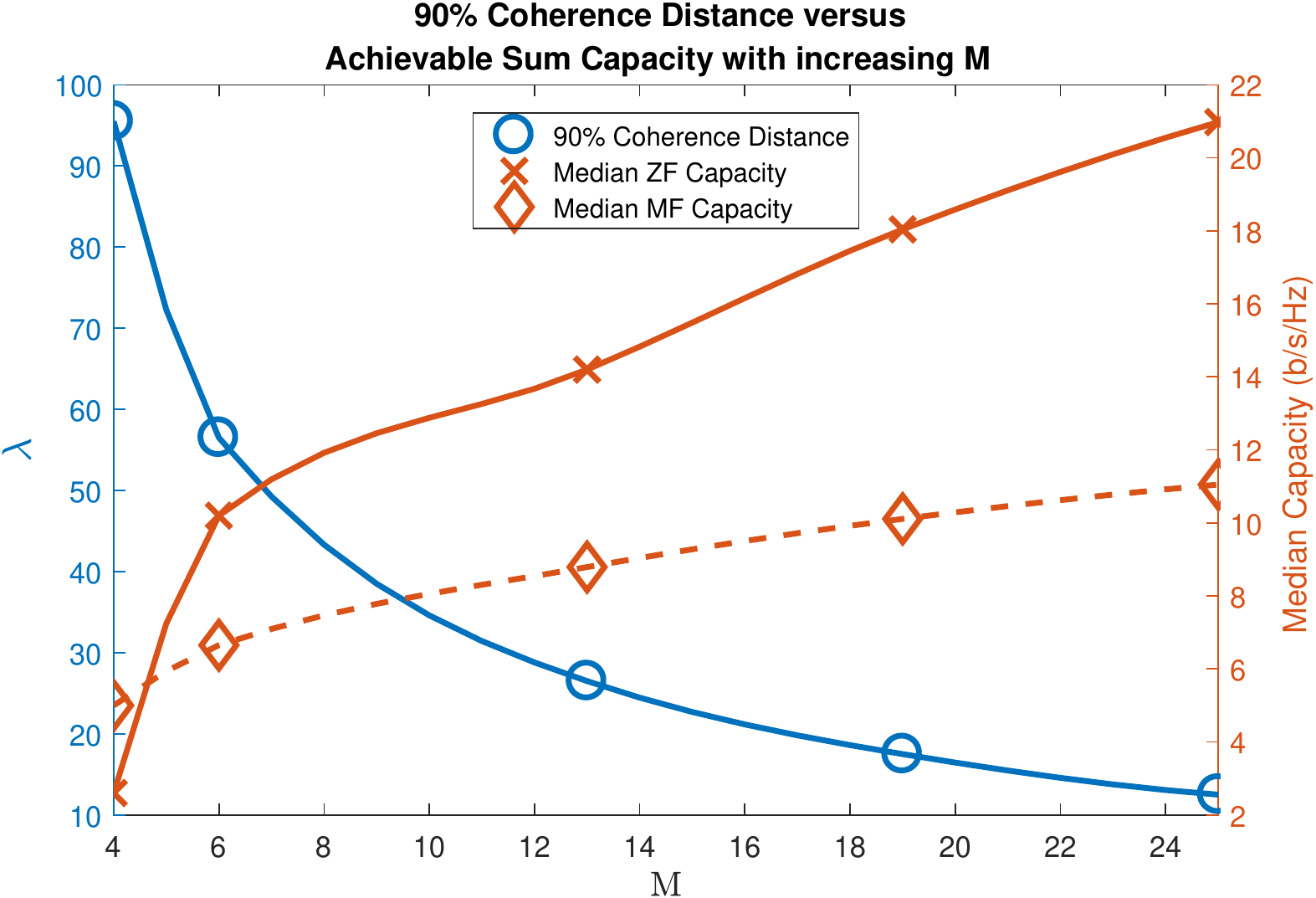}
	\caption{Terminal distance moved in wavelengths for 10\% spatial decorrelation versus median achievable sum-rate. $M$ represents the number of elements in the azimuth for the $4\times M$ array.}
	\label{fig:d10vsCapvsM}
\end{figure}

By further inspecting the temporal correlation results obtained in~\cite{PaulJSAC}, some insight can be gained on the trend of decorrelation with relation to the number of \gls{BS} antennas and terminal distance moved. As previously shown in~\cite{PaulJSAC}, the elevation dimension did not impact the rate of decorrelation in this \gls{LOS} scenario; it was primarily dictated by the azimuth resolution. Thus,~\figurename~\ref{fig:d10vsCapvsM} shows the terminal distance moved for a spatial channel decorrelation of \SI{10}{\percent} as a function of the number of \gls{BS} antennas in the azimuth only. For comparison purposes, the median achievable \gls{UL} sum-rate results have been plotted on a second axis, and a cubic interpolation has also been applied to the data points for both so that the trend can be visualized. For the achievable \gls{UL} sum-rate, the number of antennas $M$ refers to the azimuth dimension, but all four rows of elevation were used for \gls{MIMO} decoding. As the number of \gls{BS} antennas grow, it is clear that a trade-off must be made. On the one hand, increasing the number of antennas creates a greater level of pairwise orthogonality between the user channel vectors and thus increases the achievable \gls{UL} sum-rate, but the corresponding reduction in coherence time also increases the channel training requirements for each user further explaining the observations in~\cite{PaulJSAC}. 

\subsection{Expected Rate}
\begin{figure}[!t]
	\centering
	\includegraphics[width=\columnwidth]{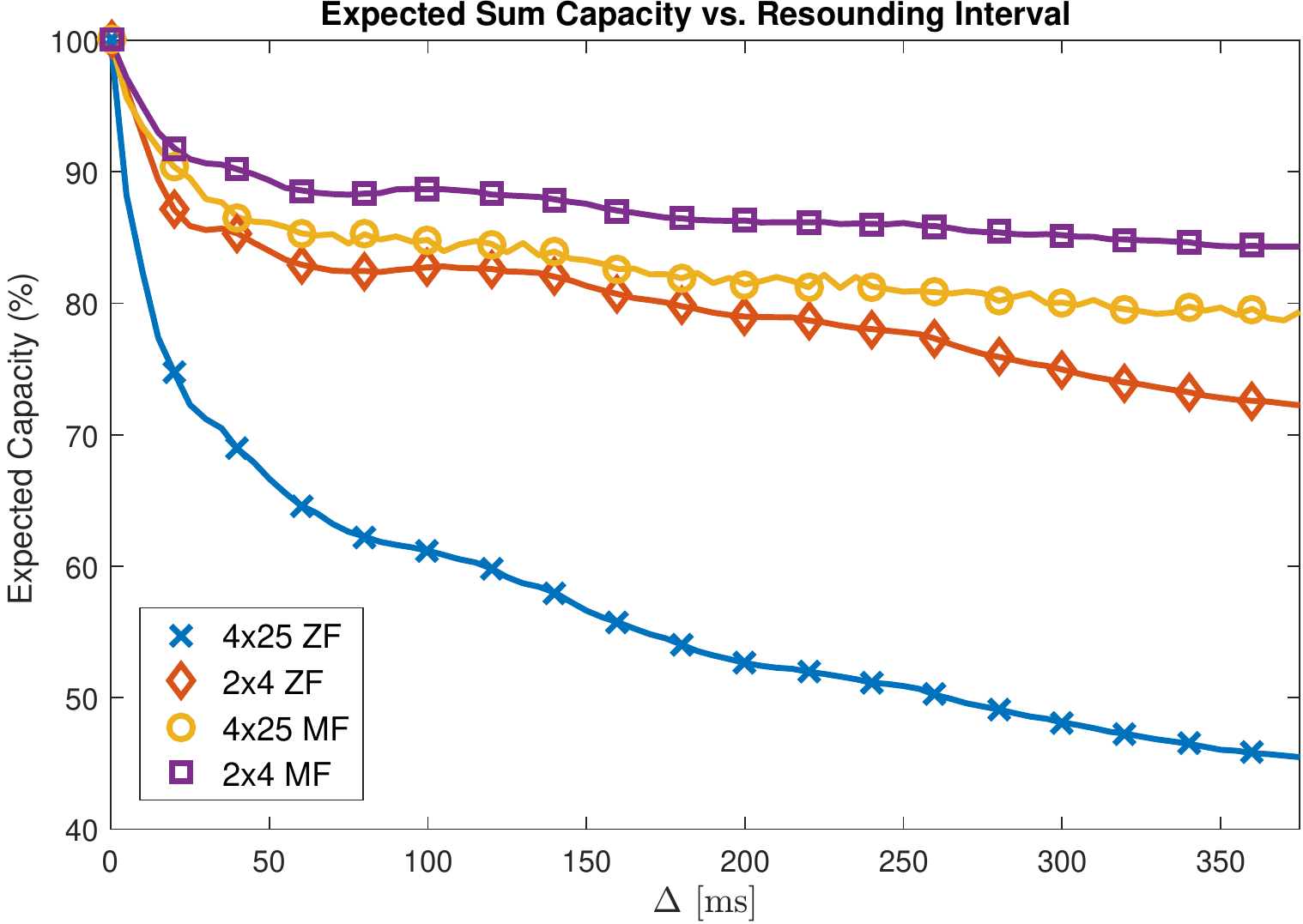}
	\caption{Expected sum-rate for \gls{ZF} and \gls{MF} versus channel resounding interval $\Delta$ across the full \SI{30}{\second} mobile scenario}
	\label{fig:ExpCap}
\end{figure}

In~\figurename~\ref{fig:ExpCap}, the expected sum-rate is shown for the \num{2x4} and \num{4x25} array configurations using both \gls{ZF} and \gls{MF}. The sensitivity of achievable performance to channel estimation error for \gls{MF} is not as severe as for \gls{ZF}, but the lack of interference mitigation typically results in a lower sum-rate when the user channel vectors are highly correlated. For \num{8} antennas, increasing $\Delta$ up to \SI{20}{\milli\second} results in an initial drop in the expected sum-rate of approximately \SI{10}{\percent}, but the performance reduction than plateaus. For the \num{100} antenna case this plateau occurs \SI{5}{\percent} lower, but the sum-rate reduction trend is very similar. This indicates the impact of channel aging on the achievable \gls{MF} sum-rate is limited in this case, even with large $M$.
\gls{ZF} performance suffers more from channel estimation error as it results in the inaccurate placement of nulls. Thus, the interference is not mitigated as effectively and the \gls{SINR} reduces for each user terminal. As with large $M$ the nulls become more focused, the sensitivity of \gls{ZF} performance to user mobility worsens, and it may be necessary to use a higher correlation percentage when calculating a suitable coherence time. The impact of $M$ upon \gls{ZF} performance with an increasing resounding interval is clearly shown in~\figurename~\ref{fig:ExpCap}. It can be seen that the exponential decay of expected sum-rate after increasing $\Delta$ by only tens of milliseconds is far more extreme for the \num{4x25} case, losing \SI{25}{\percent} of the sum-rate for $\Delta=$\SI{20}{\milli\second}. This equates to a displacement in wavelength of only 2$\lambda$ (\SI{16.2}{\centi\metre}). 

\subsection{Case Study on Outdoor LOS Training Overheads}
\begin{figure}[!t]
	\centering
	\includegraphics[width=\columnwidth]{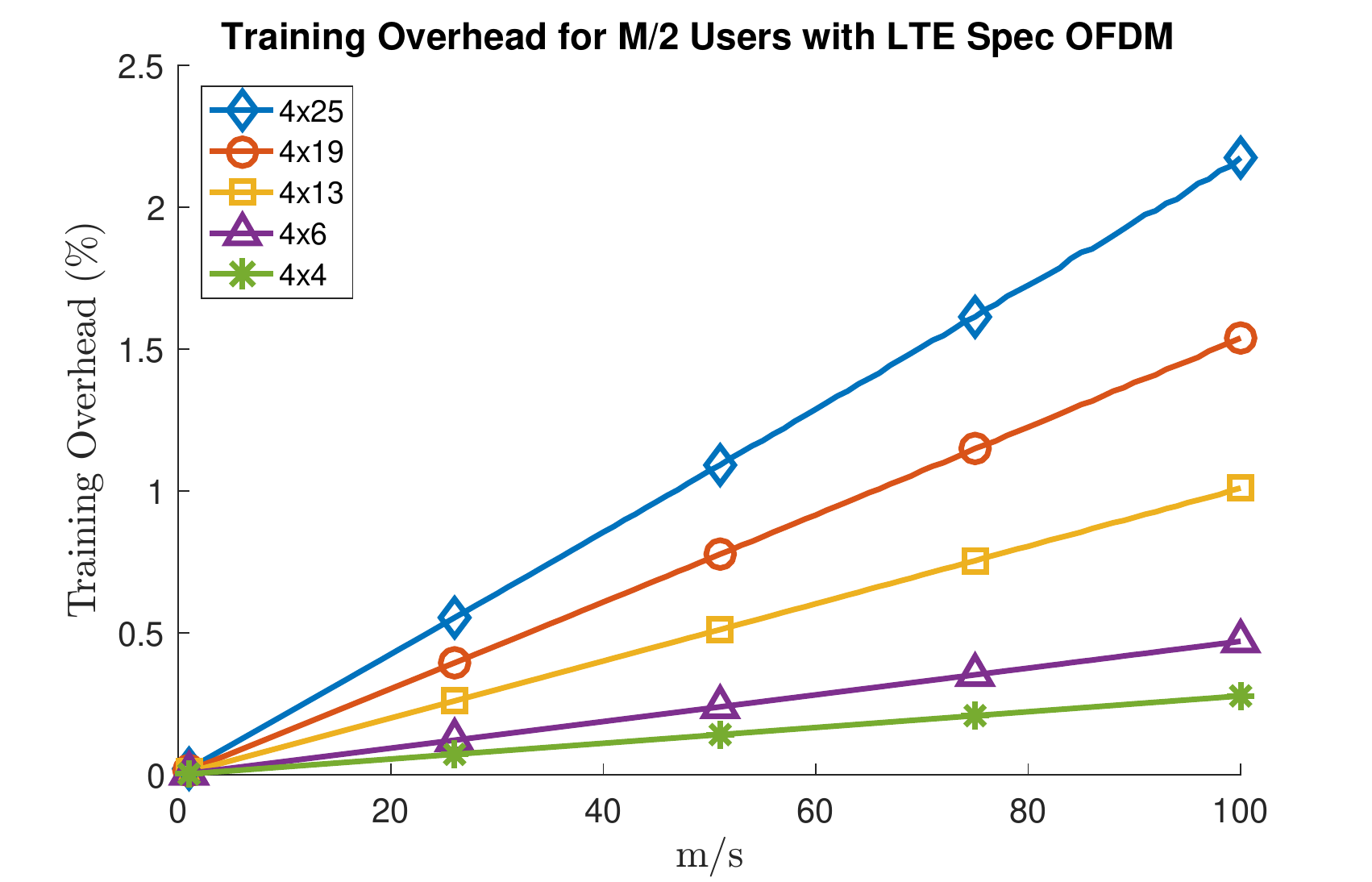}
	\caption{The percentage of OFDM symbols that will be required to train $M/2$ $(50)$ users at different speeds in the aforementioned LOS scenario. Array configurations of varying azimuth size considered using \SI{10}{\percent} decorrelation as a threshold for $T_{c}$, \SI{300}{\kilo\hertz} as a first order approximation of $B_{c}$ and an LTE OFDM symbol duration of \SI{71.4}{\micro\second}}
	\label{fig:Overhead}
\end{figure}

A case study was formed to evaluate the impact the aforementioned outdoor \gls{LOS} channel coherence results would have upon training overhead in an \gls{LTE}-spec \gls{OFDM} \gls{MMIMO} system. As channel prediction techniques such as those in~\cite{DBLP:journals/corr/abs-1305-6151} are not included here, it should be considered a worst case. The coherence interval, $\tau_{c}$, defines the number of samples available within a time and frequency coherent period for pilot and data transmission. In extreme cases where $\tau_{c}$ is very small, the number of users it is possible to train will be severely reduced and the training overhead will reduce the available samples, limiting data transmission. To formulate this case study, the background definitions for \gls{OFDM} coherence intervals and first-order coherence bandwidth approximations described in~\cite{marzetta_larsson_yang_ngo_2016} are called upon. The coherence bandwidth $B_{c}$ can be expressed in terms of \gls{OFDM} subcarriers as a frequency smoothness interval $N_{smooth}$, defined as
\begin{equation}
N_{smooth}=\left\lfloor\frac{B_{c}}{B_{s}}\right\rfloor
\end{equation}
where $B_{s}$ is the subcarrier bandwidth. Assuming a first order approximation for $B_{c}$ of \SI{300}{\kilo\hertz}, appropriate for outdoor environments, and the \SI{15}{\kilo\hertz} subcarrier width of \gls{LTE}, $N_{smooth}$ is \num{20}. Similarly, $T_{c}$ can be expressed as a number of \gls{OFDM} symbols in a coherence slot as
\begin{equation}
N_{slot}=\left\lfloor\frac{T_{c}}{T_{s}}\right\rfloor
\end{equation}
where $T_{s}$ is the \gls{OFDM} symbol duration and $T_{c}$ is the channel coherence time. For an \gls{LTE} \gls{OFDM} system, $T_{s}$ is typically \SI{71.4}{\micro\second}, and this is the value used here. $T_{c}$ was measured in~\cite{PaulJSAC} for this \gls{LOS} scenario to be \SI{125}{\milli\second} for \SI{90}{\percent} coherence at \SI{29}{\kilo\metre\per\hour}. Therefore, at an upper extreme of \SI{100}{\metre\per\second}, $T_{c}$ would become approximately \SI{10}{\milli\second} for \SI{90}{\percent} coherence, and the value for $N_{slot}$ and $\tau_{c}$ would become \num{140} and \num{2800} respectively. As $\tau_{c}>>M$ in this high mobility case, it is assumed that it will always be possible to train $M/2$ users in this study. Theoretically, this is the optimum number of users to serve when $\tau_{c}>>M$ and the \gls{SNR} is high as described in \cite{marzetta_larsson_yang_ngo_2016}. It is also assumed that all users in the system will use the full \SI{20}{\mega\hertz} system bandwidth corresponding to \num{1200} subcarriers. With an $N_{smooth}$ of 20, one \gls{OFDM} pilot symbol permits \num{20} users to obtain full \gls{CSI}. Thus, the number of symbols required for training $N_{train}$ is fixed at \num{3} and the training overhead in \SI{}{\percent} can be expressed as
\begin{equation}
T_{OH} = \frac{N_{train}}{N_{slot}} \times 100
\end{equation}
~\figurename~\ref{fig:Overhead} shows the channel training overhead $T_{OH}$ against terminal velocity for the \gls{LOS} scenario studied here using different array sizes. It shows that even in the most extreme case using \num{100} antennas with the terminal moving at \SI{100}{\metre\per\second} in parallel to the azimuth of the \gls{BS} array, typical of a high-speed train scenario, the channel training overhead is just a little over \SI{2}{\percent}. With more realistic vehicular speeds of up to \SI{30}{\metre\per\second}, it is around \SI{0.5}{\percent}. These results indicate that outdoor environments with a strong \gls{LOS} characteristic can be highly stable for \gls{MMIMO} systems and a far larger azimuth dimension could be permitted in this case. By increasing the azimuth dimension and thereby the \gls{LOS} angular resolution, spatial selectivity would be further improved and more users could be served effectively in the same time-frequency resource. $T_{c}$ would also reduce, but the above results suggest that there could be a large margin in this type of scenario, and it could be desirable to increase the training percentage of the coherence interval to obtain the \gls{MMIMO} benefits and increase the number of connected devices in the spatial domain. As an example, this would bode well for the long-range rural applications Facebook are targeting with Project Aries in \cite{Facebook}, as the 96-antenna system has a \num{2x48} azimuth dominated configuration. Finally, when following the \gls{MMIMO} system design process shown in~\cite{Malkowsky2016b}, the improved channel stability encountered within these \gls{LOS} scenarios would relax the \gls{TDD} \gls{MMIMO} precoding turnaround constraints and reduce the cost of commercial system development.

\section{Conclusions}
\label{sec:conc}
This paper has presented results for the achievable \gls{MMIMO} sum-rate, the impact of channel aging and the required training overheads for a \num{100}-antenna \gls{MMIMO} \gls{OFDM} system in a measured \gls{LOS} scenario with mobility. The results serve as accompanying material for the work originally presented in~\cite{PaulJSAC}. When serving \num{8} single antenna user terminals, it was shown that a median \gls{UL} sum-rate of \SI{20}{\bits\per\second\per\hertz} and \SI{10}{\bits\per\second\per\hertz} could be achieved for \gls{ZF} and \gls{MF} respectively. Furthermore, \SI{25}{\percent} of the achievable \gls{UL} \gls{ZF} sum-rate was lost by increasing the resounding interval by a factor of \num{4}. The channel de-coherence in that case was measured to be just \SI{3}{\percent}, indicating how sensitive \gls{ZF} can be to channel aging. Finally, the training overhead for the \gls{LOS} scenario considered assuming an \gls{LTE}-like \gls{OFDM} physical layer was calculated to be \SI{2}{\percent} at \SI{90}{\metre\per\second}. This implies a large margin could be available in such scenarios for a \gls{MMIMO} system to train more users, support greater terminal velocity and relax the precoding turn-around latency constraints in a \gls{TDD} configuration.


%

\appendices


\section*{Acknowledgment}
The authors acknowledge and thank all academic staff and post graduate students involved at both Lund University and the University of Bristol who contributed to the measurement trial operations. They also acknowledge the financial support of the \gls{EPSRC} \gls{CDT} in Communications (EP/I028153/1), 
NEC and \gls{NI}. 

\ifCLASSOPTIONcaptionsoff
  \newpage
\fi



\bibliographystyle{MyIEEEtran}
\bibliography{thesisbiblio}

\begin{thebibliography}{1}
\providecommand{\url}[1]{#1}
\csname url@samestyle\endcsname
\providecommand{\newblock}{\relax}
\providecommand{\bibinfo}[2]{#2}
\providecommand{\BIBentrySTDinterwordspacing}{\spaceskip=0pt\relax}
\providecommand{\BIBentryALTinterwordstretchfactor}{4}
\providecommand{\BIBentryALTinterwordspacing}{\spaceskip=\fontdimen2\font plus
\BIBentryALTinterwordstretchfactor\fontdimen3\font minus
  \fontdimen4\font\relax}
\providecommand{\BIBforeignlanguage}[2]{{%
\expandafter\ifx\csname l@#1\endcsname\relax
\typeout{** WARNING: IEEEtran.bst: No hyphenation pattern has been}%
\typeout{** loaded for the language `#1'. Using the pattern for}%
\typeout{** the default language instead.}%
\else
\language=\csname l@#1\endcsname
\fi
#2}}
\providecommand{\BIBdecl}{\relax}
\BIBdecl

\bibitem{Marzetta2010}
T.~L. Marzetta, ``Noncooperative cellular wireless with unlimited Numbers of
  base station antennas,'' \emph{IEEE Transactions on Wireless Communications},
  vol.~9, no.~11, pp. 3590--3600, Nov 2010.

\bibitem{DBLP:journals/corr/abs-1305-6151}
\BIBentryALTinterwordspacing
K.~T. Truong and R.~W.~H. Jr., ``Effects of Channel Aging in Massive {MIMO}
  Systems,'' \emph{CoRR}, vol. abs/1305.6151, 2013. [Online]. Available:
  \url{http://arxiv.org/abs/1305.6151}
\BIBentrySTDinterwordspacing

\bibitem{7510708}
P.~Kela, X.~Gelabert, J.~Turkka, M.~Costa, K.~Heiska, K.~Lepp{\"a}nen, and
  C.~Qvarfordt, ``Supporting mobility in 5G: A comparison between massive MIMO
  and continuous ultra dense networks,'' in \emph{2016 IEEE International
  Conference on Communications (ICC)}, May 2016, {Kuala Lumpur, Malaysia}.

\bibitem{7869082}
C.~Shepard, J.~Ding, R.~E. Guerra, and L.~Zhong, ``{Understanding real
  many-antenna MU-MIMO channels},'' in \emph{2016 50th Asilomar Conference on
  Signals, Systems and Computers}, Nov 2016, pp. 461--467.

\bibitem{PaulJSAC}
P.~Harris, S.~Malkowsky, J.~Vieira, E.~Bengtsson, F.~Tufvesson, W.~B. Hasan,
  L.~Liu, M.~Beach, S.~Armour, and O.~Edfors, ``Performance Characterization of
  a Real-Time Massive MIMO System With LOS Mobile Channels,'' \emph{IEEE
  Journal on Selected Areas in Communications}, vol.~35, no.~6, pp. 1244--1253,
  June 2017.

\bibitem{6773067}
C.~E. Shannon, ``A mathematical theory of communication,'' \emph{The Bell
  System Technical Journal}, vol.~27, no.~4, pp. 623--656, Oct 1948.

\bibitem{marzetta_larsson_yang_ngo_2016}
T.~L. Marzetta, E.~G. Larsson, H.~Yang, and H.~Q. Ngo, \emph{Fundamentals of
  Massive MIMO}.\hskip 1em plus 0.5em minus 0.4em\relax Cambridge University
  Press, 2016.

\bibitem{Facebook}
\BIBentryALTinterwordspacing
``{Facebook Project Aries},'' 2016. [Online]. Available:
  \url{https://code.facebook.com/posts/1072680049445290/introducing-facebook-s-new-terrestrial-connectivity-systems-terragraph-and-project-aries/}
\BIBentrySTDinterwordspacing

\bibitem{Malkowsky2016b}
S.~Malkowsky, J.~Vieira, L.~Liu, P.~Harris, K.~Nieman, N.~Kundargi, I.~C. Wong,
  F.~Tufvesson, V.~{\"O}wall, and O.~Edfors, ``The World's First Real-Time
  Testbed for Massive MIMO: Design, Implementation, and Validation,''
  \emph{IEEE Access}, vol.~5, pp. 9073--9088, 2017.

\end{thebibliography}
\end{document}